# Extending the operating window of scanning electron microscopy through an integrated electron-optical architecture for high-temperature and near-ambient-pressure environments

Yue Chai[1], Honglong Zhao[1], Xinning Tian[1], Chao Ang[1], Zhu-Jun Wang[1*]

**Corresponding author: Zhu-Jun Wang, wangzhj3@shanghaitech.edu.cn*

1. School of Physical Science and Technology, ShanghaiTech University, Shanghai 201210, China.

**ABSTRACT**

Scanning electron microscopy (SEM) under simultaneously high-temperature, near-ambient-pressure (NAP), and reactive-gas environments requires coordinated control of vacuum isolation, electron-beam transmission, signal generation, and thermal management—constraints that have long limited the operating window of environmental scanning electron microscopy (ESEM). Here we establish an integrated electron-optical architecture that combines a multi-stage differential pressure pathway, front-stage pressure transition, detector optimization, thermal management, and a gas-focusing sampling architecture into a unified ESEM platform. Pressure distribution and electron-beam transmission are quantitatively validated through computational fluid dynamics (CFD), Monte Carlo electron–gas scattering analysis, and direct beam-current measurements, while detector optimization and thermionic-electron suppression preserve stable imaging under elevated pressure and temperature. The resulting system enables stable SEM imaging at pressures up to 20,000 Pa and high-temperature imaging up to 1,400 °C under 6,000 Pa of air, continuous observation of hydrated biological specimens, and synchronized SEM–QMS operando characterization using localized gas sampling. These developments establish a general electron-optical framework for extending ESEM toward realistic *operando* environments where elevated temperature, reactive gases, structural evolution, and gas-phase chemistry can be investigated simultaneously.

**KEYWORDS:**

Scanning electron microscopy, Electron-optical architecture, Differential pumping, Electron-beam transport, NAP microscopy, High-temperature in situ imaging

## INTRODUCTION

ESEM has transformed conventional SEM from a vacuum-limited characterization technique[1,2] into a versatile platform capable of imaging specimens in controlled gaseous environments. By integrating differential pumping with gaseous electron detection, commercial ESEM systems enable direct observation of insulating, hydrated, and environmentally sensitive specimens without requiring high-vacuum conditions[3-8], substantially expanding the applicability of SEM across materials science, catalysis, geology, and biological research. As a mature and widely adopted technology, ESEM has become an indispensable tool for investigating dynamic processes that cannot be studied using conventional SEM alone.

Increasingly, however, many scientifically and technologically important processes occur under conditions that combine elevated temperatures with NAP reactive-gas environments. Representative examples include catalytic reactions[9-13], crystal growth and dynamic structural evolution [7,14-16], oxidation and corrosion[14,17-20], mineral transformations[14,21], and biological processes under hydrated conditions[6,22]. Performing electron microscopy under these realistic environments would enable direct visualization of dynamic structural evolution while maintaining the chemical and thermodynamic conditions under which these processes naturally occur. Extending the operating window of ESEM toward simultaneously high-temperature and NAP environments has therefore become an important objective for next-generation ESEM.

Despite the success of ESEM, achieving this operating regime remains a fundamental instrumentation challenge because the entire electron-optical pathway becomes strongly coupled under elevated pressure and temperature. Increasing gas pressure substantially enhances electron–gas scattering, reducing the effective incident beam current and degrading spatial resolution, while the electron source must still be protected by ultrahigh vacuum (UHV) through efficient differential pumping[3,23,24]. Elevated temperatures further introduce thermal radiation, thermionic electron backgrounds, mechanical drift, and increased gas loading, which collectively affect signal generation, detector stability, and long-term system operation[14,25,26]. Consequently, vacuum isolation, electron-beam transport, signal detection, thermal management, and environmental control can no longer be optimized independently but must instead be considered as an integrated electron-optical pathway.

Over the past decades, substantial progress has been made in addressing individual aspects of ESEM. Differential pumping systems have enabled higher chamber pressures[23,24], gaseous detectors have improved imaging under low-vacuum conditions[23], atmospheric-pressure SEM has demonstrated imaging at ambient pressure[27,28], and specialized heating stages have extended ESEM toward elevated temperatures[14,26,29,30]. These developments have significantly broadened the capability of ESEM and established the technological foundation of modern ESEM instrumentation. Nevertheless, these advances have largely optimized individual components or operating conditions independently. As summarized in Fig. 1, reported systems generally achieve either high-temperature imaging at relatively low pressures or high-pressure imaging near room temperature, whereas simultaneous operation under high-temperature, NAP, and reactive-gas environments (such as $H_2$, CO, $NO_2$) remains largely inaccessible[24,26-28,30]. This fragmented operating window reflects the absence of a unified electron-optical architecture capable of simultaneously coordinating pressure management, beam transport, signal detection, thermal control, and environmental compatibility.

Here, we establish an integrated electron-optical architecture that extends the operating window of ESEM toward high-temperature and NAP operation. Rather than introducing a single instrumental component, the platform systematically redesigns the electron-optical pathway from the electron source to the specimen environment. A multi-stage differential vacuum system incorporating a millimeter-scale front-stage differential pumping module (DPM) establishes a stable pressure gradient while minimizing electron–gas scattering within the critical high-pressure region. Electron-beam transmission is quantitatively characterized through CFD simulations, Monte Carlo electron–gas scattering analysis, and direct beam-current measurements using a newly developed beam-separating Faraday cup (BSFC). To maintain imaging performance under elevated pressure and temperature, the platform further integrates a gas-compatible electron-ionization detector (EID) featuring a large-area detection geometry and integrated thermal management, enabling operation at an ultrashort working distance (WD) to minimize electron–gas scattering while maintaining detector stability. The platform also incorporates localized laser-confined heating, thermionic-electron suppression structures, and synchronized quadrupole mass spectrometry (QMS). Together, these developments provide a unified instrumentation framework that substantially extends the operating window of ESEM toward realistic operando environments.

## RESULTS AND DISCUSSION

### Electron-optical architecture for extending the operating window of ESEM

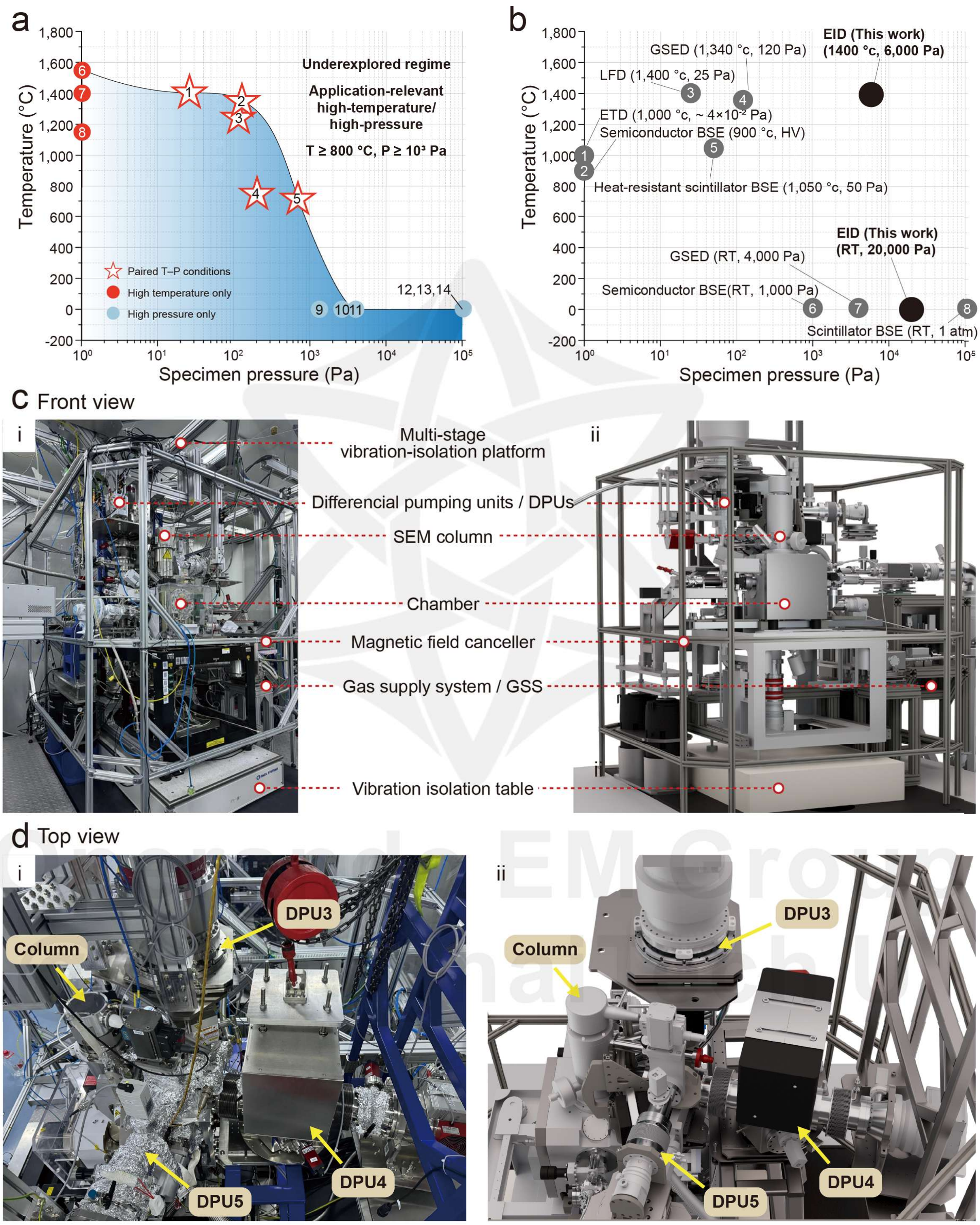

**FIG. 1.** Design concept and overall configuration of the integrated ESEM platform. **(a)** Temperature–pressure operating window of previously reported SEM systems, highlighting the limited overlap between high-temperature and NAP operation. **(b)** Representative detector strategies and their corresponding representative operating temperature–pressure conditions reported in the literature. The EID conditions at 1,400 °C and 6,000 Pa and at room temperature and 20,000 Pa were obtained in separate experiments and do not represent a single paired operating point or a continuous operating envelope. **(c)** Overall configuration of the integrated SEM platform. (i) Photograph of the complete instrument showing the vibration-isolation platform, differential pumping units (DPUs), SEM column, specimen chamber, magnetic-field cancellation system, gas supply system (GSS), and vibration-isolation table. (ii) Three-dimensional rendering of the complete platform illustrating the spatial arrangement of the major system components. **(d)** Core microscope configuration. (i) Top-view photograph of the SEM column, chamber, and differential pumping modules. (ii) corresponding three-dimensional rendering highlighting the spatial relationship between the SEM column and the multi-stage differential pumping system.

As discussed above, extending the operating window of ESEM toward simultaneously high-temperature and NAP operation requires coordinated optimization of the entire electron-optical pathway rather than isolated improvements to individual components[14,23,24,26]. We therefore established an integrated electron-optical architecture that systematically couples pressure management, electron-beam transport, signal detection, thermal management, and environmental control into a unified instrumentation platform.

To define the operating range targeted by this architecture, Fig. 1a,b summarizes the temperature–pressure conditions reported for SEM-based environmental imaging. Previous studies have demonstrated high-temperature in situ imaging, kilopascal-pressure imaging, and atmospheric-pressure imaging under different experimental configurations (see Figs. S1 and S2 and Tables S1 and S2 in the supplementary material)[7,12,16,17,21,22,24,26-34]. High-temperature imaging has generally been performed at relatively low pressures, whereas imaging at kilopascal to atmospheric pressures has mostly been achieved at room temperature or moderately elevated temperatures. Representative examples include atmospheric-pressure SEM using pressure-limiting structures with backscattered electron (BSE) detection and high-temperature ESEM using a gaseous secondary electron detector (GSED) at approximately 120 Pa and temperatures approaching 1,340 °C[27,29]. The detector configurations summarized in Fig. 1b therefore occupy distinct temperature–pressure ranges[24,26,27,30,35].

The limited overlap between high-temperature and high-pressure operation arises from the combined constraints of detector operation, electron–gas scattering, and thermal management. Scintillator BSE detectors can operate at elevated pressures, but high-temperature imaging requires heat-resistant designs and thermal management to reduce radiative heating and thermal load[26,27]. The associated increase in WD or gas path length enhances electron–gas scattering and reduces the effective signal[23,24]. GSEDs are well suited to elevated pressures and can also be adapted for high-temperature imaging[24,30]. At high temperature, however, thermionic electrons from the specimen increase the detection background, while thermal management constrains the detector–specimen geometry[14,30]. These coupled effects limit simultaneous operation at high temperature and near-ambient pressure.

To overcome these coupled limitations, the platform was organized around a continuous electron-optical pathway extending from the electron source to the specimen environment (Fig. 1c,d). The system integrates a multi-stage differential vacuum system, gas supply and pressure-control modules, localized laser heating with associated thermal-management measures, signal detection, and environmental control into a unified architecture designed specifically for operation under elevated pressure and temperature. In addition, the platform incorporates multi-stage vibration isolation and an external magnetic-field cancellation system to suppress mechanical and electromagnetic disturbances during long-term operation. Figures 1c and 1d illustrate the overall system architecture and the spatial organization of its principal functional modules, including the SEM column, specimen chamber, DPUs, GSS, and supporting environmental-control components.

Within this architecture, maintaining a continuous electron-optical pathway while simultaneously preserving ultrahigh vacuum in the electron column constitutes the fundamental prerequisite for stable operation under NAP conditions[3,23,24]. Consequently, pressure management forms the first design objective of the integrated electron-optical architecture, providing the vacuum boundary upon which subsequent electron-beam transport, signal generation, and high-temperature imaging all depend. The following section therefore first establishes the multi-stage differential vacuum system that defines the pressure distribution throughout the electron-optical pathway.

**Pressure management through a multi-stage differential electron-optical pathway**

Pressure management constitutes the first prerequisite for preserving electron-optical performance under NAP operation. Because the electron beam propagates through an uninterrupted electron-optical pathway, ultrahigh vacuum must be maintained in the electron column while the specimen chamber simultaneously operates under NAP gaseous environments[3,24]. Rather than relying on discrete vacuum isolation, this requirement demands a continuous pressure transition along the entire beam path. To achieve this objective, we established a multi-stage differential electron-optical pathway consisting of five vacuum chambers (VC1–VC5) and their corresponding DPUs (DPU1–DPU5), thereby generating a stable pressure gradient extending from the specimen chamber (10,000 Pa) to the electron gun ($10^{-7}$ Pa) (Fig. 2a,b). DPU1, implemented as the DPM and associated with VC1, forms the front-stage pressure-limiting structure between the specimen and the pole piece.

Within this pressure architecture, the most critical challenge occurs at the entrance of the electron-optical pathway beneath the pole piece, where the gas density is highest and electron–gas scattering is most severe. Because scattering probability increases rapidly with pressure, pressure reduction must occur within the shortest possible distance to preserve beam coherence before significant scattering accumulates[23,24]. Accordingly, the simulated pressure distribution shows that DPU1 rapidly reduces the local pressure from approximately 10,000 Pa to the order of 10 Pa within only a few millimeters, thereby suppressing beam degradation at the earliest stage of beam propagation (Fig. 2c). The subsequent differential pumping stages progressively reduce the simulated pressure to approximately $10^{-1}$ Pa (DPU2), $10^{-5}$ Pa (DPU3), and finally maintain ultrahigh vacuum (~$10^{-7}$ Pa) within the electron gun through DPU4 and DPU5. Besides protecting the electron source, the final differential pumping stages employ ion pumping together with non-evaporable getter (NEG) pumping to enhance the removal of low-molecular-weight residual gases, particularly $H_2$, thereby maintaining the required vacuum conditions in the upstream electron-optical pathway[36,37]. The pumping capacities and pressure stages were jointly optimized to maintain stable long-term operation under NAP conditions while preserving compatibility with multiple reactive-gas environments.

To evaluate whether the proposed pressure architecture satisfies these design requirements, the pressure distribution along the electron-optical pathway was quantitatively characterized using CFD simulations together with experimental pressure measurements[8,38]. CFD simulations were performed under practical operating conditions with a specimen-chamber pressure of 10,000 Pa ($N_2$) and a specimen-to-DPU1 entrance distance of 1 mm. Axial positions in Fig. 2 are referenced to the specimen surface along the electron-optical axis. The simulated pressure field demonstrates that the multi-stage differential architecture establishes a continuous pressure transition throughout the entire electron-optical pathway (Fig. 2c; see Figs. S3 and S4 and Table S3 in the supplementary material). Compared with the differential pumping configuration of a commercial ESEM, the proposed architecture produces a much steeper pressure drop immediately beneath the pole piece, reducing the pressure from 5,754 Pa to 21 Pa at the VC2 entrance, corresponding to a reduction greater than 99.6%. Consequently, the high-pressure region is effectively confined to the specimen side of the electron-optical pathway, substantially limiting gas penetration toward the electron column while maintaining ultrahigh vacuum in the electron gun.

Because local pressure gradients and confined gas flow complicate direct pressure characterization within differential pumping structures, comparison between CFD simulations and experimental measurements provides an effective means of validating the predicted pressure distribution[38]. As shown in Fig. 2c, the experimentally measured pressures agree well with the simulated results throughout the differential pathway. For example, the measured pressures at DPU4 and DPU5 differ from the simulated values by 22.46% and 20.49%, respectively, indicating that the CFD model provides a reasonable representation of the pressure distribution established along the differential electron-optical pathway. Measurements with $H_2$ showed a similar stagewise pressure-reduction trend. These results demonstrate that the multi-stage differential pathway provides a robust pressure boundary for the integrated electron-optical architecture and establishes the vacuum conditions required for stable electron-beam transport under NAP environments.

Having established the global pressure distribution throughout the electron-optical pathway, the remaining dominant limitation to beam transmission arises from localized electron scattering within the high-pressure entrance region beneath the pole piece. The following section therefore focuses on engineering this front-stage pressure transition through a compact DPM to further suppress beam degradation before significant electron–gas scattering occurs.

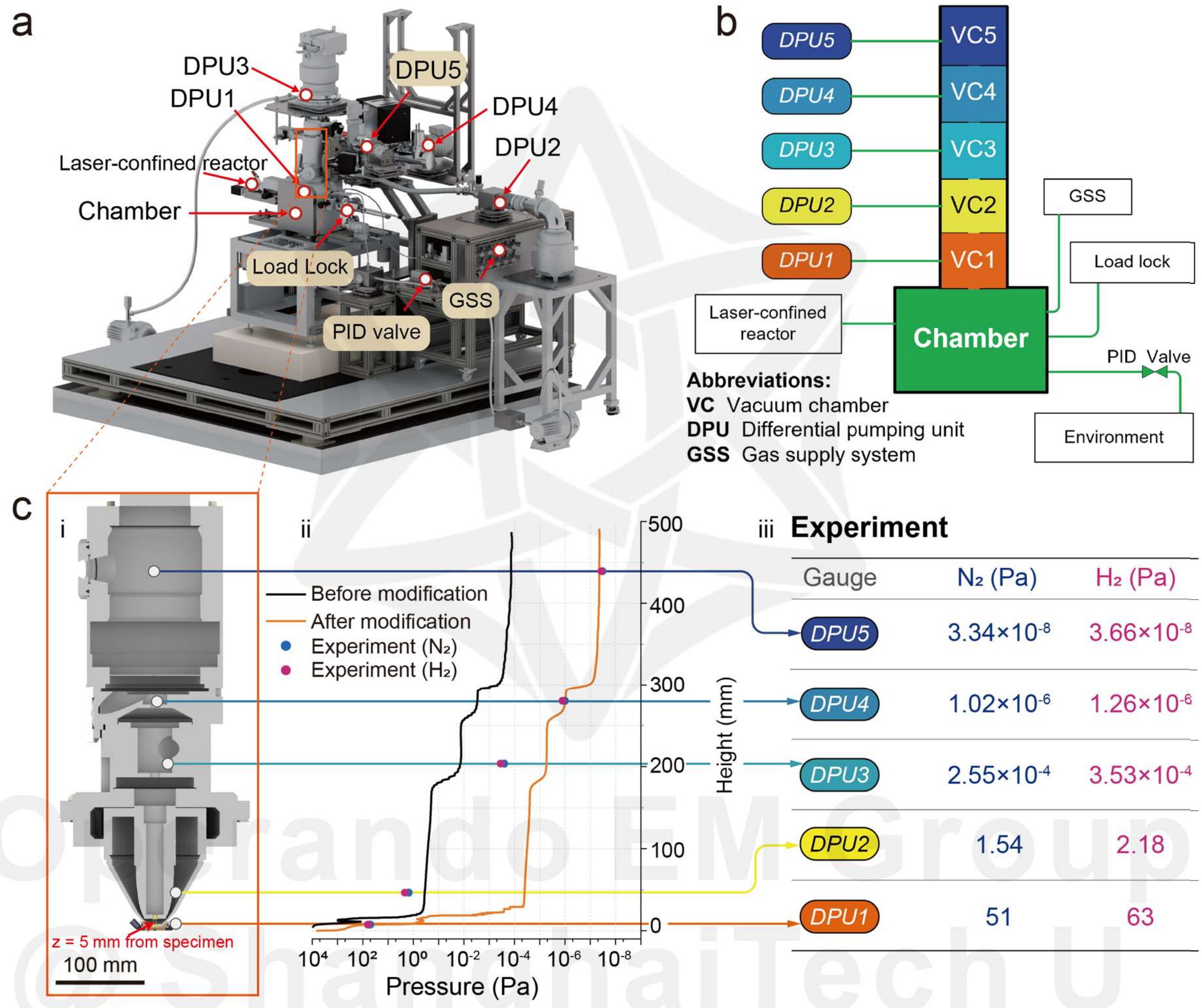

**FIG. 2.** Pressure management within the integrated electron-optical architecture. **(a)** and **(b)** Overall configuration of the multi-stage differential pressure architecture. VC1–VC5 and DPU1–DPU5 are sequentially arranged along the electron-optical pathway to establish a continuous pressure transition between the NAP specimen chamber and the ultrahigh-vacuum electron column. The specimen chamber integrates a load lock, laser-confined reactor, GSS, and proportional-integral-derivative (PID) valve for sample transfer, localized heating, gas delivery, and pressure regulation. **(c)** Quantitative validation of the pressure distribution along the differential electron-optical pathway. (i) Cross-sectional view of the electron column showing the locations of the pressure gauges. (ii) Simulated axial pressure profiles for a commercial ESEM (black) and the present architecture (orange); blue and magenta symbols indicate the experimentally measured pressures under $N_2$ and $H_2$ atmospheres, respectively. $H_2$ values are uncorrected gauge indications. (iii) Measured pressures at DPU1–DPU5 under $N_2$ and $H_2$ operating conditions.

## Engineering the front-stage pressure transition

Although the multi-stage pressure architecture establishes the global pressure distribution along the electron-optical pathway, the dominant source of beam degradation remains localized beneath the pole piece, where the incident electron beam first enters the high-pressure environment. Because electron–gas scattering increases rapidly with gas density, preserving beam coherence requires an extremely steep pressure transition within only a few millimeters before substantial scattering occurs[23,24]. Engineering this front-stage pressure transition therefore becomes the second key design objective of the integrated electron-optical architecture.

To realize such an ultracompact pressure transition, we explored a series of geometries capable of suppressing reverse gas transport while maintaining an uninterrupted electron-optical pathway. The design originated from the Tesla valve concept (Fig. 3a, stage i), which exploits geometric asymmetry to impede reverse flow[39]. However, the conventional Tesla valve is not compatible with electron transmission because its flow path is not axially continuous[40]. We therefore progressively evolved the design into a family of multi-stage differential vortex orifice array (Multi-DVOA) (Fig. 3a, stages i-v), which preserve a continuous axial beam path while introducing directional flow resistance through geometric vortex generation.

The design evolution was evaluated using CFD simulations together with scaled experimental models (Fig. 3a,b; see Fig. S5 and Table S4 in the supplementary material). In the simulations, the inlet pressure was fixed at 10,000 Pa $N_2$, while identical pumping conditions were applied to compare the pressure-reduction characteristics of the different geometries. Scaled models of Designs ii–v were fabricated by three-dimensional printing and tested under the same pumping configuration using the steady downstream pressure as the experimental metric. Among the experimentally tested geometries, Design v produced the lowest downstream pressure. The conventional Tesla-valve geometry served only as the conceptual starting point for the subsequent Multi-DVOA development.

Because the available space beneath the pole piece is severely constrained, the final bilateral-pumping Multi-DVOA geometry (Design v in Fig. 3a) cannot be directly implemented within the microscope column. Considering pressure reduction efficiency, manufacturability, and geometric compatibility, the geometries represented by Designs iv and v were therefore combined in a compact DPM (Fig. 3c,d). The DPM consists of three sequential Multi-DVOA sections. Multi-DVOA1 is located outside the pole piece and establishes the front-stage differential pumping region associated with VC1, whereas Multi-DVOA2 and Multi-DVOA3 extend through VC2 and together form an axially permeable unidirectional vortex structure (APUVS). Multi-DVOA1 and the APUVS together constitute the complete DPM. The DPM further incorporates a water-cooling tube to limit heat transfer from the high-temperature specimen region toward the pole piece and adjacent components (see Fig. S6 in the supplementary material).

According to the simulated pressure distribution, Multi-DVOA1 adopts Design v and rapidly reduces the local pressure from 10,000 Pa to below $10^2$ Pa within less than 5 mm, whereas the APUVS aperture adopts Design iv and further lowers the pressure to $4.93 \times 10^{-3}$

Pa at the outlet of VC2 through the combined action of directional vortex structures and intermediate pumping. Together, these elements establish a highly compact front-stage pressure transition that efficiently isolates VC1 from VC2 while maintaining a continuous electron-optical pathway.

The effectiveness of the front-stage pressure transition was further evaluated using local CFD simulations and Monte Carlo electron-transport analysis[41]. Compared with the original differential pumping configuration, the DPM redirects gas flow toward the lateral pumping channels, confines the high-pressure region immediately beneath the pole piece, and significantly accelerates pressure decay within VC2 (Fig. 3e,f). Based on the simulated local pressure field, the no-collision primary-beam fraction, $T$, was calculated to be approximately 2.7% after transmission through the high-pressure entrance region (Fig. 3g). Although most primary electrons undergo at least one scattering event within this region, a finite no-collision beam component remains along the electron-optical axis[42,43].

Together, these results demonstrate that the DPM effectively reshapes the local pressure field beneath the pole piece, establishing the steep front-stage pressure transition required to preserve electron transmission before substantial electron–gas scattering occurs. Having engineered the local pressure architecture, the following section quantitatively evaluates its impact on electron-beam preservation through direct measurements of the effective incident beam current.

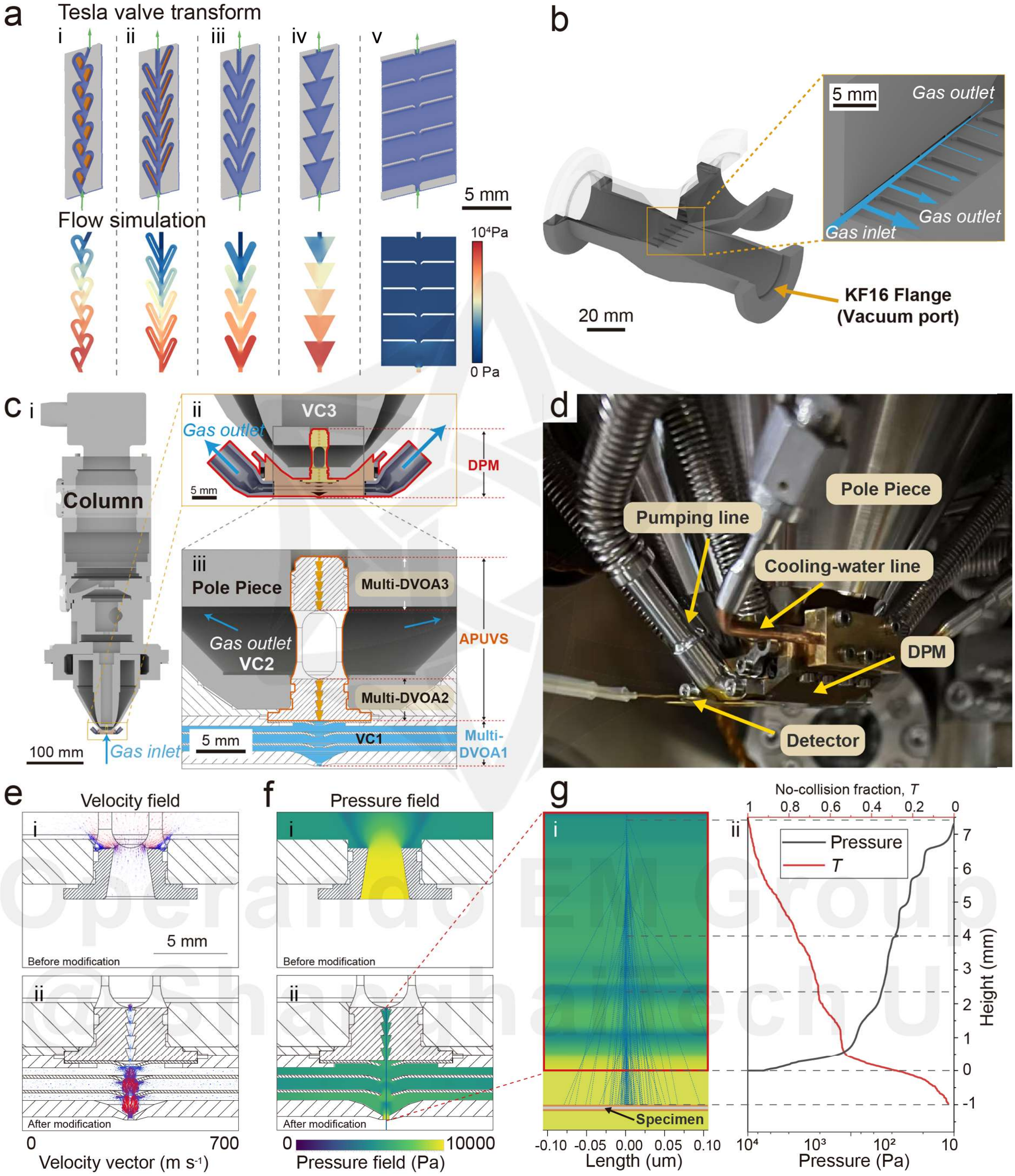

**FIG. 3.** Engineering the front-stage pressure transition beneath the pole piece. **(a)** Evolution of the flow-directing geometry from a conventional Tesla valve to Multi-DVOA geometries, together with the corresponding CFD flow simulations. **(b)** Scaled rendering of Design v illustrating the evaluated geometry. **(c)** and **(d)** Configuration and implementation of the DPM. **(e)** and **(f)** Comparison of gas velocity and pressure fields between the original differential pumping configuration and the DPM architecture. **(g)** Monte Carlo simulation of electron-beam transport through the locally engineered pressure field, showing electron trajectories together with the pressure distribution and no-collision primary-beam fraction along the electron-optical pathway.

**Quantitative characterization of electron-beam transport**

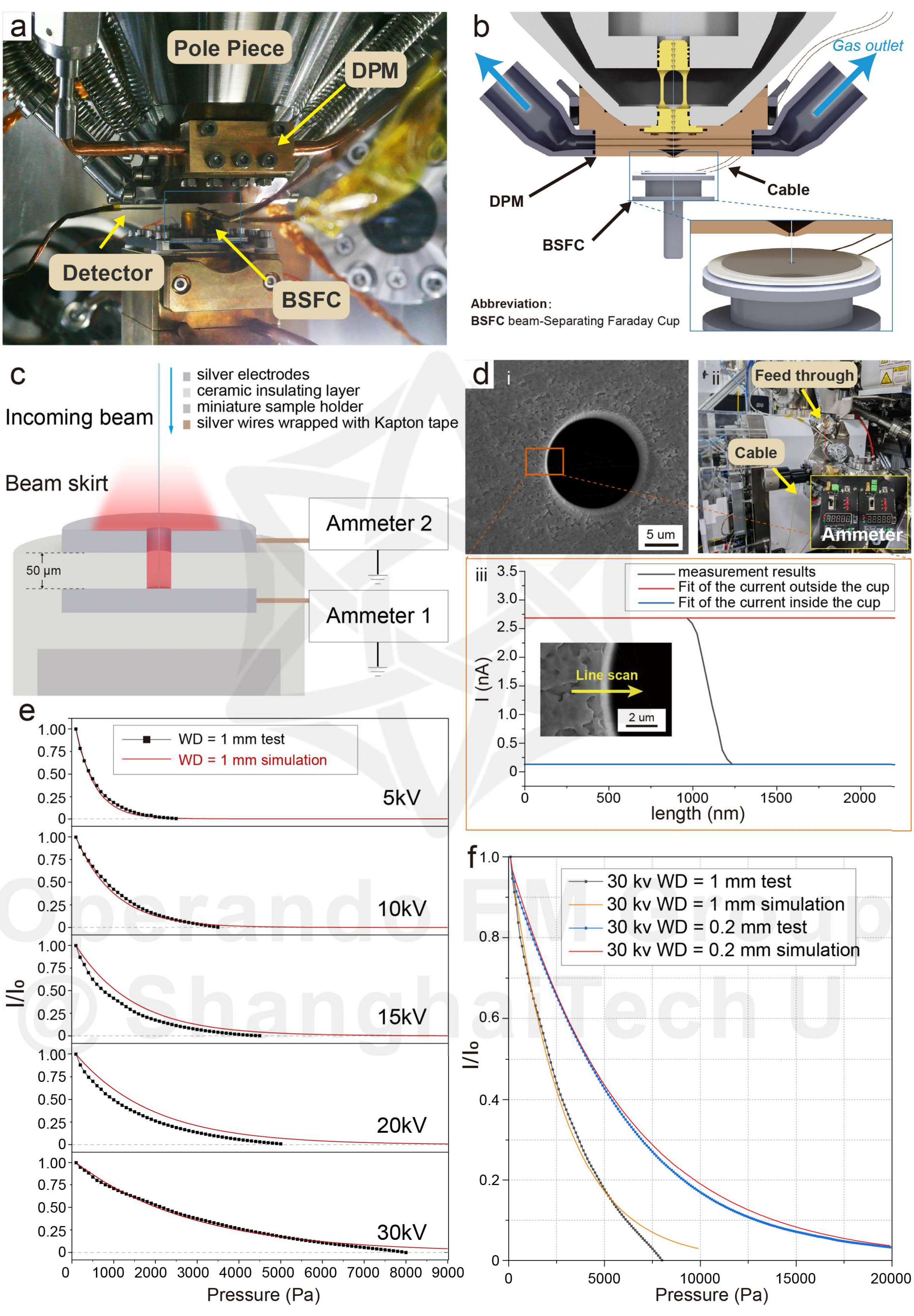

**FIG. 4.** Quantitative characterization of electron-beam transport under NAP conditions. **(a)** and **(b)** Implementation of the BSFC beneath the DPM. **(c)** Working principle of the BSFC for spatially separating the central-aperture and peripheral electron-current components according to electron arrival position. **(d)** Experimental verification of the spatial-separation capability using a transverse electron-beam scan across the central aperture. **(e)** Pressure-dependent comparison of the calculated and BSFC-derived beam-transport metrics, each normalized to its respective value at 100 Pa and expressed as $I/I_0$. **(f)** Effect of working distance on the two normalized beam-transport metrics at an accelerating voltage of 30 kV.

The effectiveness of the proposed pressure architecture ultimately depends on whether the central primary-beam component can be maintained under NAP conditions. As the incident beam propagates through the gaseous environment, electron–gas scattering redistributes primary electrons from the central beam into a surrounding beam skirt, thereby reducing the central current density and broadening the spatial distribution of electrons at the specimen plane[23,24,42]. Quantitative characterization of this pressure-dependent central-beam attenuation is therefore required to evaluate the proposed electron-optical architecture.

To characterize the spatial redistribution of the electron beam experimentally, a BSFC was integrated immediately beneath the DPM (Fig. 4a–c)[42,44]. The BSFC consists of two coaxial silver electrodes separated by a 50 μm insulating layer. The upper annular electrode contains a central aperture of approximately 10 μm, and the lower collecting electrode is positioned directly beneath the aperture. Electrons arriving within the central-aperture region pass to the lower electrode, whereas electrons arriving outside this region are intercepted by the upper annular electrode. Independent electrical connections allow the currents collected by the two electrodes to be recorded simultaneously. The BSFC therefore separates electron-current components according to their arrival positions rather than according to the scattering histories of individual electrons.

Before the pressure-dependent measurements, the spatial-separation capability of the BSFC was verified using a transverse electron-beam scan across the central aperture, following established Faraday-cup beam-profile evaluation approaches[24,45]. During the scan, the currents from the upper annular and lower collecting electrodes were recorded simultaneously. The two currents exhibited complementary variations as the beam crossed the aperture boundary (Fig. 4d), confirming the spatial discrimination between electrons passing through the central aperture and those intercepted by the upper annular electrode. The line scan was also used to determine the beam-alignment position for the subsequent pressure-dependent measurements. This measurement verifies spatial separation by electron arrival position but does not determine whether an individual electron has undergone gas scattering.

Electron-beam transport was subsequently evaluated using Monte Carlo calculations and direct BSFC measurements[41]. In the calculation, the transmitted component is defined by the no-collision fraction $T(p)$, whereas the BSFC derives a central-beam current from aperture-edge differential measurements. Small-angle scattered electrons may still reach the BSFC collection region, whereas any electron undergoing a scattering event is excluded from the calculated no-collision population. The calculated and measured quantities therefore describe related beam-transport behavior but retain different operational definitions.

Based on the pressure distribution obtained from the CFD calculations (Fig. 3), the high-pressure region extending from Multi-DVOA1 to the specimen plane was selected as the electron–gas interaction region for Monte Carlo modeling. Increasing gas pressure within this region increases electron–gas scattering and attenuates the central beam component[23,24]. The normalized BSFC current, $I/I_0$, was defined relative to the aperture-edge differential current measured at 100 Pa. For consistency in graphical comparison, the calculated no-collision fraction was likewise expressed as $I/I_0 = T(p)/T(100\text{ Pa})$. Thus, $I_0$ denotes the respective 100 Pa reference value for each method, rather than an identical physical quantity. Further details are provided in Secs. S6 and S7 of the supplementary material.

As shown in Fig. 4e, both the measured and calculated $I/I_0$ values decrease with increasing chamber pressure. Increasing the accelerating voltage reduces the pressure-dependent attenuation of both quantities, consistent with the reduced electron–gas scattering probability of higher-energy electrons under the investigated conditions[41,42,46]. The Monte Carlo and BSFC results exhibit similar pressure- and voltage-dependent trends, supporting the ability of the scattering calculation to capture the principal attenuation behavior of the central beam component. Quantitative equality between the two quantities is not expected because they are defined using different criteria.

At an accelerating voltage of 30 kV, reducing the working distance from 1.0 mm to 0.2 mm decreases the length of the near-specimen high-pressure gas path and thereby increases both normalized beam-transport metrics at a given chamber pressure (Fig. 4f). Under the shorter-working-distance condition, the BSFC-derived central-beam signal remained measurable above 10,000 Pa. These results show that increasing the accelerating voltage and shortening the high-pressure propagation path reduce the loss of the central beam component under NAP conditions.

Together, the Monte Carlo calculations and BSFC measurements provide complementary quantitative characterization of electron-beam transport under NAP operation. The results establish how chamber pressure, accelerating voltage, and working distance affect preservation of the central beam component. The following section therefore focuses on converting the electrons reaching the specimen region into stable imaging signals under elevated-pressure conditions.

## Detector optimization under NAP conditions

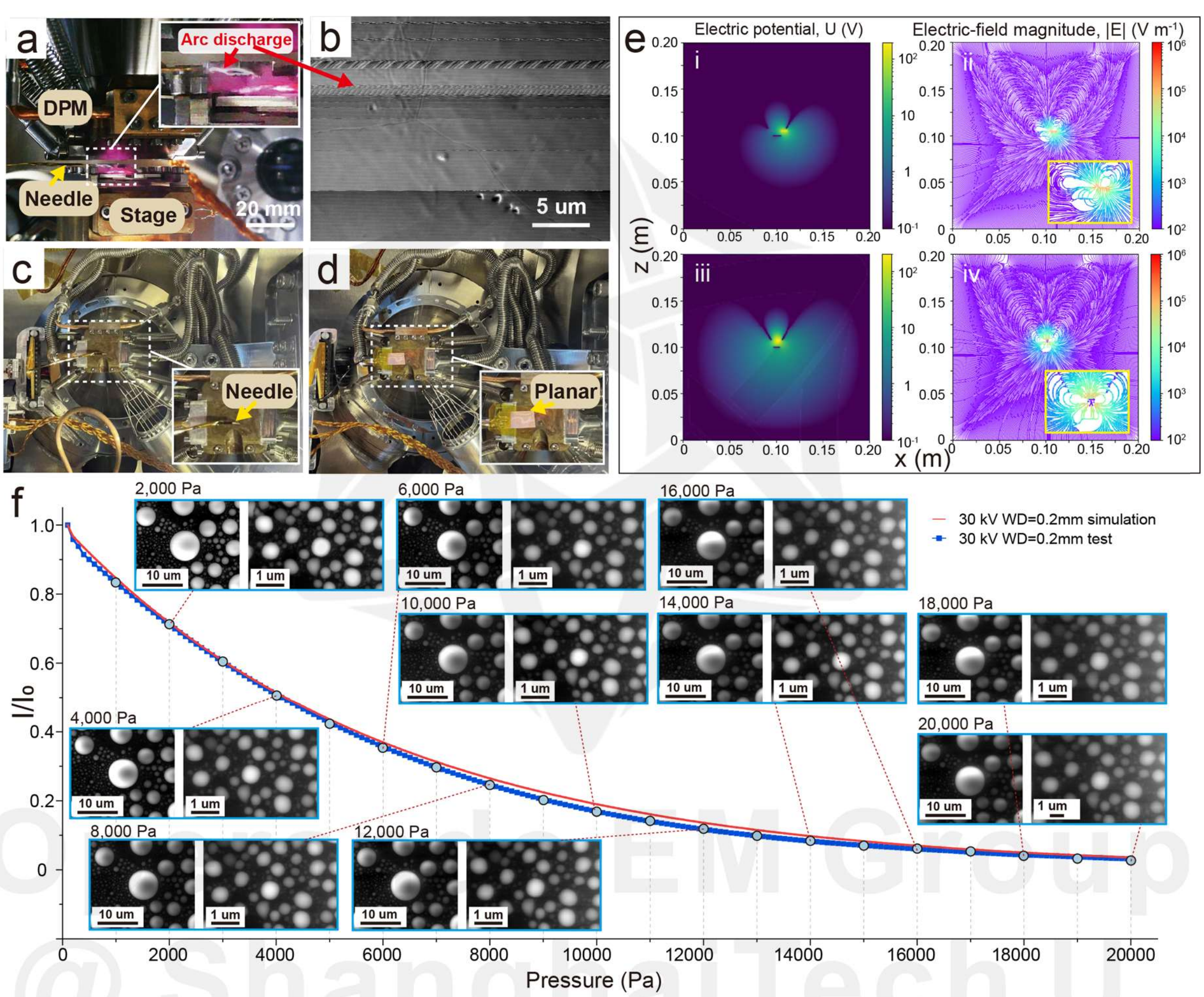

**FIG. 5.** Detector optimization for stable signal generation under NAP conditions. **(a)** and **(b)** Arc-discharge limitation of the needle-electrode detector under elevated-pressure conditions. **(c)** and **(d)** Detector redesign from a needle electrode to a planar electrode for improved electrical stability. **(e)** Simulated electric-potential and electric-field distributions of the two detector configurations. **(f)** Pressure-dependent normalized beam-transport metrics obtained from the BSFC measurements and calculations. Insets, EID images of Sn spheres acquired in $N_2$ at 2,000–20,000 Pa, 30 kV, 0.31 nA, and WD = 0.2 mm. The EID bias was manually adjusted for stable signal collection at each pressure point; the images are therefore used as a qualitative pressure-series demonstration.

Although electron transmission is preserved, stable imaging further depends on efficient signal generation under NAP conditions. Unlike conventional high-vacuum SEM, conventional ESEM commonly employs gas-amplified secondary-electron detection, in which secondary electrons emitted from the specimen initiate gas ionization and subsequent signal amplification[4,5]. At sufficiently high chamber pressures, the reduced electric field under a given detector geometry and bias becomes insufficient to sustain efficient collision ionization, thereby limiting secondary-electron gas amplification[24,47]. Consequently, stable signal generation becomes another critical requirement for imaging under elevated-pressure conditions.

To overcome this limitation, Rattenberger et al. proposed introducing a needle electrode near the specimen to generate a highly localized electric field capable of accelerating secondary electrons beyond the gas-ionization threshold[47]. Following this strategy, we first implemented a needle electrode with a tip size of approximately 10 μm between the DPM and the specimen (Fig. 5a). Although this configuration effectively enhances gas amplification, experiments revealed that the highly concentrated electric field readily induces arc discharge at elevated chamber pressures (Fig. 5a,b), leading to abrupt intensity fluctuations and image instability. These observations indicate that maximizing local electric-field strength alone is insufficient for reliable imaging under NAP conditions.

The detector design was therefore reformulated with a different objective: rather than maximizing local electric-field intensity, the goal was to achieve stable signal generation through controlled electric-field distribution. To this end, the needle electrode was replaced by a planar electrode with a thickness of 50 μm (Fig. 5c,d). The planar geometry effectively homogenizes the electric field, thereby suppressing local field enhancement and reducing the probability of electrical breakdown while maintaining efficient collection of gas-amplified electrons. To further minimize electron scattering during signal collection, the detector was positioned directly above the specimen, reducing both the WD and the electron collection path. A 50 μm Kapton insulating layer was incorporated to electrically isolate the detector from the DPM while introducing only a minimal increase in WD (see Sec. S8 in the supplementary material). The optimized detector architecture is hereafter referred to as the EID.

To compare the electric-field distributions produced by the two detector geometries, three-dimensional electrostatic calculations were performed (Fig. 5e). The needle electrode generates a highly concentrated electric field near its tip, enabling rapid acceleration of secondary electrons and efficient gas ionization, but simultaneously producing localized field maxima that promote electrical breakdown under elevated-pressure conditions. In contrast, the planar electrode distributes the electric field much more uniformly, substantially reducing peak field strength while maintaining an effective electron-collection region. Combined with its shorter collection distance and larger collection area, the planar geometry reduces localized field concentration and provides a broader collection region. The electrostatic calculation is used only to compare field geometry and does not directly predict gas gain or collection efficiency.

The EID was finally evaluated under progressively increasing chamber pressure using Sn spheres as a standard imaging specimen (Fig. 5f). Imaging experiments were performed from 2,000 to 20,000 Pa while maintaining identical specimen position, beam current, and accelerating voltage. Although image contrast gradually decreases with increasing chamber pressure, the detector maintains stable imaging performance throughout the investigated pressure range. For reference, the pressure-dependent attenuation of the central primary-beam component obtained from the BSFC measurements and Monte Carlo calculations is also shown in Fig. 5f. The inset SEM images demonstrate that clear Sn-sphere morphology remains observable with the EID up to 20,000 Pa. These images qualitatively demonstrate EID imaging from 2,000 to 20,000 Pa after manual bias optimization at each pressure condition.

Together, these results demonstrate that the EID successfully completes the electron-optical pathway from transmitted primary electrons to stable image formation under NAP conditions. Having established stable signal generation, the remaining challenge for extending the operating window of SEM is to suppress thermally generated electron backgrounds during high-temperature operation. The following section therefore focuses on thermal management and high-temperature imaging.

# Thermal management for high-temperature imaging

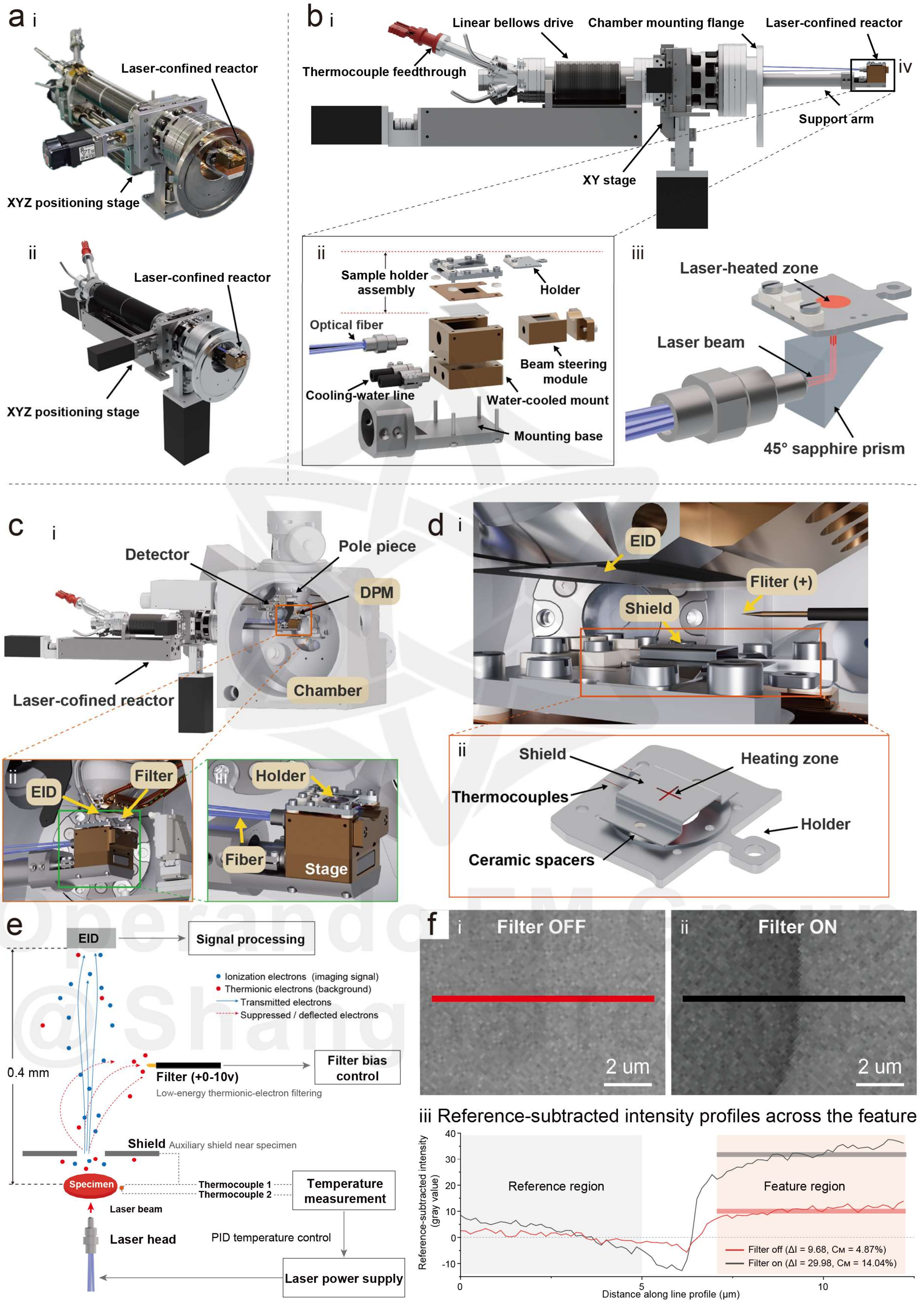

**FIG. 6.** Thermal management for stable high-temperature imaging. **(a)** and **(b)** Localized laser-heating architecture and reactor configuration. **(c)** Integration of the laser-confined reactor into the electron-optical architecture together with the DPM, EID, and specimen stage. **(d)** and **(e)** Thermionic-electron suppression strategy using the combined shield–filter configuration and closed-loop temperature control. **(f)** Experimental validation of thermionic-electron background suppression. Images acquired with the filter switched off and on, together with quantitative comparison of the reference-subtracted intensity profiles demonstrating enhanced image contrast during high-temperature operation.

Elevated temperature introduces two coupled challenges to NAP imaging: increased thermal loading of the electron-optical components and thermionic-electron interference with the gas-amplified imaging signal. The former can lead to excessive heating of the pole piece and detector region, whereas the latter introduces an additional electron background originating directly from the heated specimen[14,25]. Consequently, both thermal-load management and thermionic-electron suppression are required for stable high-temperature imaging.

To achieve localized high-temperature operation while minimizing thermal loading of the surrounding chamber components, a laser-confined reactor was developed (Fig. 6a,b). The reactor employs a linear bellows drive together with an XY translation stage to provide precise positioning of the heating assembly. Three laser beams are delivered through optical fibers and redirected by a 45° sapphire prism to produce localized heating directly beneath the specimen. Compared with conventional resistive heating, which typically heats a much larger thermal mass surrounding the specimen, focused laser heating confines the high-temperature region to the immediate vicinity of the specimen, thereby reducing the heated volume and the associated thermal load on the surrounding microscope components[48,49]. Water cooling integrated into the reactor further removes excess heat and limits heat transfer to adjacent chamber components[14].

Thermal management was also incorporated directly into the detector-side electron-optical structure. Under NAP conditions, reducing the detector–specimen separation is essential for minimizing the electron–gas interaction path during signal collection and maximizing the detection efficiency of gas-amplified secondary electrons. This requirement becomes even more critical at elevated chamber pressures, where electron scattering rapidly increases with propagation distance. However, positioning the detector close to a high-temperature specimen inevitably increases the thermal load on the detector assembly, presenting a major limitation for conventional ESEM detector configurations[14]. To overcome this trade-off, the EID was mounted directly onto the water-cooled DPM, which functions not only as the front-stage differential pumping structure but also as an actively cooled thermal sink. The water-cooled DPM provides an active heat-removal path for heat transferred toward the detector assembly, while the thin planar geometry of the EID limits its thermal mass. Together, these features facilitate thermal management of the detector assembly and allow operation at a short detector–specimen separation during high-temperature experiments.

Although active cooling limits the thermal load on the detector-side electron-optical components, thermionic electrons emitted from the heated specimen remain a distinct source of imaging background[14,26]. To reduce this interference, a positively biased surface-sensitive filter was introduced between the specimen and the EID to deflect low-energy thermionic electrons before they enter the signal-collection region[50]. For experiments above approximately 1000 °C, an additional shield was positioned close to the specimen to geometrically restrict the direct propagation of thermionic electrons toward the detector region (Fig. 6d,e)[29]. The shield and filter therefore provide complementary suppression mechanisms: the shield intercepts part of the thermionic-electron flux close to the specimen, whereas the positively biased filter further modifies the contribution of residual low-energy electrons to the detected signal before gas amplification. Specimen temperature is continuously monitored by a thermocouple and regulated through a PID control loop (see Secs. S9 and S10 and Table S5 in the supplementary material).

The effectiveness of thermionic-electron background suppression was evaluated by comparing images acquired with the filter switched on and off (Fig. 6f). With the filter activated, image contrast increased, consistent with a reduction or redistribution of the thermionic-electron contribution to the detected signal. Quantitative analysis of the reference-subtracted intensity profiles shows that the feature intensity difference increases from 9.68 to 29.98, while the Michelson contrast increases from 4.87% to 14.04%. These improvements are consistent with the expected suppression of low-energy thermionic electrons and demonstrate that thermionic-electron management significantly enhances image quality during high-temperature operation.

Collectively, these results demonstrate that stable high-temperature imaging relies on the combined control of thermal load and thermionic-electron background. Localized laser heating confines the high-temperature region to the vicinity of the specimen, while the water-cooled DPM and thermally managed EID preserve an ultrashort detector–specimen separation without compromising detector stability. The thermionic-electron shield and surface-sensitive filter further suppress background electrons generated at elevated temperatures. Together with closed-loop temperature regulation, these developments extend the operating window of the proposed electron-optical architecture from NAP imaging toward simultaneous high-temperature and NAP operation.

**System-level validation under representative operating environments**

Having established pressure management, electron-beam transport, signal generation, and thermal management, the complete electron-optical architecture was finally evaluated under representative operating environments that simultaneously challenge multiple aspects of system performance. To evaluate the platform under distinct operating conditions, two complementary classes of specimens were selected: a high-temperature non-conductive ceramic precursor system representing reactive gaseous environments, and hydrated living biological specimens representing NAP imaging under high-humidity conditions. Together, these demonstrations evaluate the operating window enabled by the proposed electron-optical architecture.

The first demonstration used a non-conductive system under high-temperature gaseous conditions. A high-melting insulating glass-forming system was selected to demonstrate stable SEM imaging under simultaneous high-temperature and near-ambient-pressure conditions. The $BaSO_4$–$SiO_2$ precursor mixture exhibits severe charging under conventional high-vacuum SEM (Fig. 7a), making continuous observation of its high-temperature morphological evolution difficult. To evaluate the capability of the present platform, *in situ* heating experiments were performed under 6,000 Pa of air while the specimen temperature was monitored using a Type B thermocouple (Fig. 7b; see Fig. S8a and Sec. S11 in the supplementary material). $BaSO_4$ has a reported melting point of approximately 1580 °C[51], providing a suitable model system for evaluating stable SEM imaging close to the melting regime of insulating glass-forming materials. As the temperature increased to 1063 °C, the specimen remained predominantly solid. Localized surface softening and the onset of melting became apparent at approximately 1100 °C. With further heating, the molten regions progressively expanded, and by 1400 °C a continuous liquid phase had developed across a substantial portion of the specimen while stable SEM imaging was maintained throughout the entire transformation. These observations demonstrate that the proposed platform enables continuous visualization of melting and liquid-phase evolution in an insulating glass-forming system under simultaneous high-temperature and near-ambient-pressure conditions.

The second demonstration evaluates the capability of the platform for hydrated biological specimens under NAP conditions. Living mites were first imaged at a total pressure of 20,000 Pa while the water-vapor partial pressure was gradually reduced (Fig. 7c,d). Under the initial hydrated condition, the overall body morphology, surface features, and appendages were clearly resolved, and locomotion of the appendages could be continuously observed. As the water-vapor partial pressure decreased, the movement gradually weakened and eventually ceased at the lowest water-vapor partial pressure investigated (100 Pa), while the total pressure remained at 20,000 Pa. Mobility did not recover within the recorded observation period under the present experimental conditions. In a second experiment, both the total pressure and water-vapor partial pressure were simultaneously reduced to monitor morphological evolution observed during simultaneous reductions in total and water-vapor pressures (Fig. 7e; see Fig. S8b and Tables S6 and S7 in the supplementary material). Progressive dehydration resulted in obvious shrinkage and wrinkling of the body surface, whereas a separate dead mite imaged under high-vacuum conditions exhibited severe charging artifacts. Together, these experiments demonstrate continuous observation of both locomotor activity and dehydration-induced structural evolution under controlled environmental conditions.

Together, these demonstrations verify that the proposed electron-optical architecture maintains stable imaging performance across two representative operating environments: high-temperature observation of non-conductive ceramic materials under a gaseous environment and NAP observation of hydrated biological specimens. Rather than validating individual instrumental components, these experiments demonstrate that the complete system successfully extends the operating window of environmental scanning electron microscopy toward practical in situ characterization under complex environmental conditions.

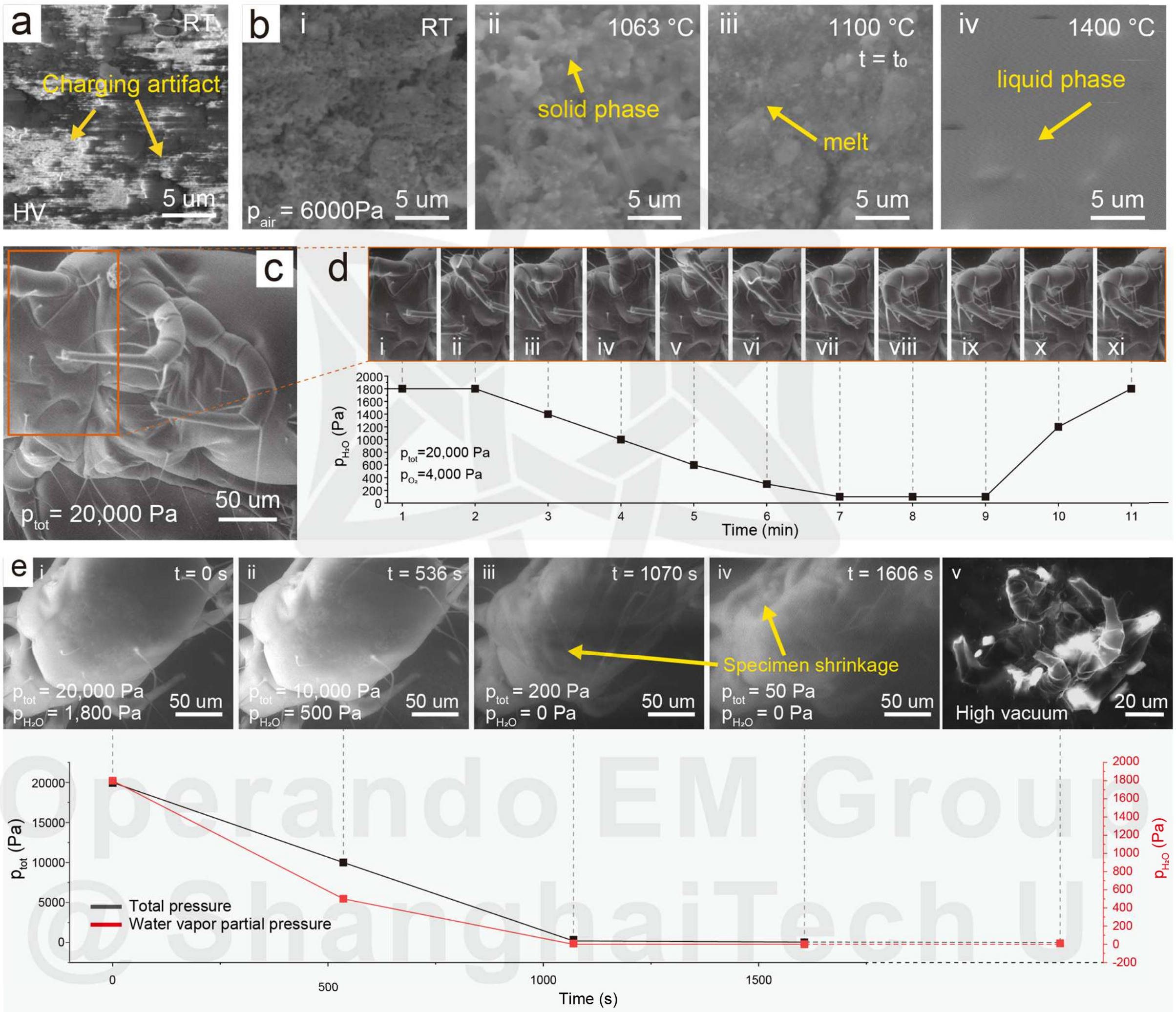

**FIG. 7.** Demonstration of the electron-optical architecture under representative operating environments. **(a)** Charging of a non-conductive $BaSO_4$–$SiO_2$ precursor under high-vacuum conditions. **(b)** High-temperature morphological evolution of the precursor under 6,000 Pa of air. **(c)** and **(d)** NAP imaging of living mites during reduction of the water-vapor partial pressure, showing changes in appendage motion. **(e)** (i)–(iv) Morphological evolution of a living mite during simultaneous reduction of the total pressure and water-vapor partial pressure, showing progressive dehydration-induced shrinkage and surface wrinkling; (v) high-vacuum SEM image of an independently prepared dead mite, included as a reference for charging under conventional high-vacuum imaging.

## Gas-sampling optimization for synchronized SEM–QMS operando characterization

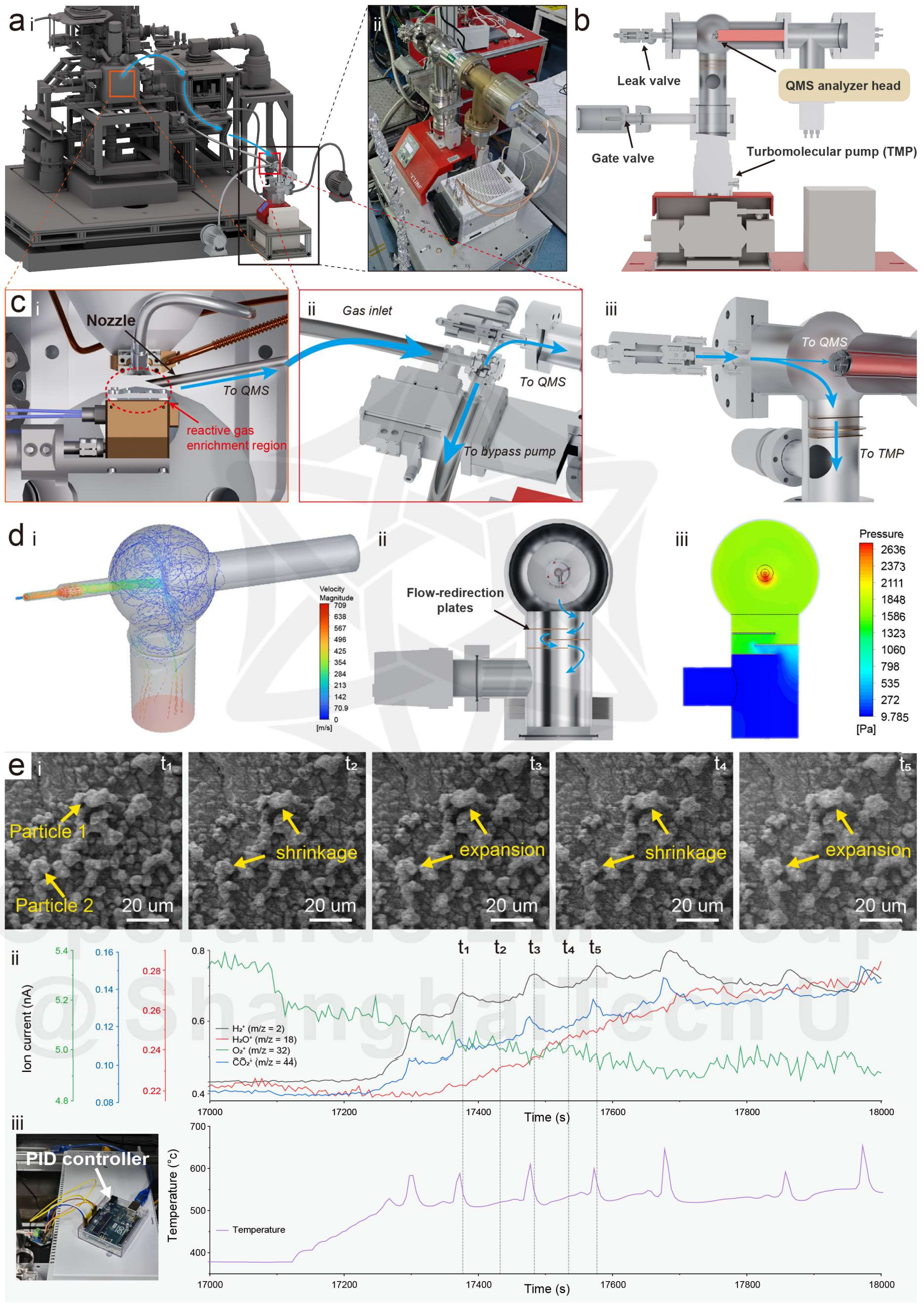

**FIG. 8.** Localized gas-sampling interface for synchronized SEM–QMS operando measurements. **(a)** and **(b)** Overall configuration of the high-sensitivity mass spectrometric sampling interface integrated with the SEM–QMS platform. **(c)** Local reaction-zone sampling interface illustrating representative gas extraction from the specimen region. **(d)** Flow and pressure simulations of the gas-focusing sampling architecture. Flow-field distributions and pressure-field simulations demonstrate directed gas transport toward the QMS analyzer-head inlet and pressure redistribution across the flow-directing structure. **(e)** Operando demonstration during the cyclic morphological evolution of Ni particles, showing the qualitative temporal alignment of SEM images, specimen temperature, and relative QMS ion-current signals after a nominal transport-delay correction.

While environmental imaging provides direct structural information, operando studies additionally require synchronized measurement of the surrounding gas-phase chemistry[10,11,52]. Achieving this capability requires representative sampling of reaction gases directly from the local reaction region while preserving the environmental conditions established for imaging. Consequently, the principal instrumentation challenge is not the mass spectrometer itself, but the development of a gas-sampling architecture capable of simultaneously providing high analytical sensitivity, rapid gas exchange, and pressure isolation.

To address this challenge, a dedicated gas-focusing sampling architecture was developed between the specimen chamber and the QMS (Fig. 8a,b). The QMS was connected to the specimen chamber through a local sampling nozzle positioned adjacent to the reaction region. Sampled gases are continuously extracted from the vicinity of the specimen and transported through a dedicated sampling pathway toward the analyzer. The QMS analyzer-head inlet opens into a spherical upstream mixing chamber, while a precision leak valve and turbomolecular pump maintain the vacuum required for mass spectrometric analysis. Rather than simply coupling SEM with QMS, the sampling architecture is specifically designed to facilitate gas transport toward the analyzer-head inlet of locally generated reaction products while minimizing disturbance to the specimen environment.

The gas-sampling pathway is illustrated in Fig. 8c. Reaction gases are first extracted through a nozzle positioned immediately above the specimen surface and transported toward the leak valve. Rapid transport of sampled gases to the mass spectrometer is important for maintaining temporal correspondence between structural and gas-phase measurements[53]. Upstream of the leak valve, a bypass branch continuously removes the majority of the sampled gas, enabling rapid gas renewal within the sampling line, while only a small controlled fraction enters the spherical mixing chamber before reaching the QMS analyzer head. This configuration simultaneously maintains representative sampling of the local reaction environment, rapid gas renewal, and pressure isolation between the specimen chamber and the mass spectrometer.

To further enhance analytical sensitivity, the gas-focusing mechanism was quantitatively evaluated using coupled flow-field and pressure-field simulations (Fig. 8d). Flow-field simulations show that gas entering through the leak valve is preferentially directed toward the QMS analyzer-head inlet rather than dispersing uniformly throughout the chamber. To suppress backflow from the vacuum-gauge/pumping branch and minimize its influence on the QMS signal, a three-stage flow-directing structure was introduced between the analyzer-head inlet region and the vacuum-gauge/pumping branch. The corresponding pressure-field simulations, performed with $N_2$ and an atmospheric-pressure boundary imposed at the leak-valve inlet, show that the elevated pressure remains confined near the analyzer head inlet and decreases rapidly downstream of the flow-directing structure, indicating effective pressure isolation from the downstream vacuum-gauge/pumping branch. Together, these results demonstrate that the proposed gas-focusing sampling architecture directs sampled gas toward the analyzer-head inlet while maintaining effective pressure isolation. The integrated SEM–QMS capability was demonstrated using the oscillatory redox of Ni particles in a $C_2H_4/O_2$ atmosphere (Fig. 8e), consistent with restructuring behavior reported for Ni catalysts in related reaction environments[54]. During cyclic temperature variation, periodic shrinkage and expansion of the Ni particles were continuously recorded by SEM while QMS simultaneously monitored the evolution of $H_2$, $H_2O$, $O_2$, and $CO_2$. By synchronizing SEM imaging, QMS analysis, and temperature measurement on a common time axis, structural evolution could be directly correlated with changes in the relative ion-current signals at the monitored $m/z$ channels throughout the reaction process (see Sec. S12 and Table S8 in the supplementary material). The local sampling configuration enabled reliable detection of transient gas-phase species while maintaining uninterrupted imaging of the reacting catalyst.

Together, these results demonstrate that the proposed gas-focusing sampling architecture extends the electron-optical architecture beyond environmental imaging toward synchronized operando characterization. By integrating a localized gas-focusing sampling architecture, high-sensitivity mass spectrometry, in situ electron microscopy, and temperature measurement into a unified platform, the system enables direct correlation between structural evolution, gas-phase chemistry, and thermal evolution under realistic reaction conditions, thereby extending the analytical capability of the platform from environmental imaging to synchronized operando characterization.

## CONCLUSIONS

This work establishes an integrated electron-optical architecture that systematically extends the operating window of ESEM toward simultaneous high-temperature, NAP, and reactive-gas environments. By coordinating pressure management, electron- beam transport, signal generation, thermal management, and operando integration within a unified instrumentation framework, the platform overcomes the coupled instrumental limitations that have traditionally restricted ESEM under realistic operating conditions.

The proposed architecture establishes a stable pressure boundary spanning multiple orders of magnitude between the specimen chamber and the electron source, while the compact DPM reshapes the local pressure field to preserve electron transmission through the high-pressure entrance region. The pressure-field validation, together with the similar pressure- and voltage-dependent trends obtained from the no-collision calculation and BSFC measurements, supports the relationship between the engineered pressure architecture and preservation of the central beam component, providing an experimental basis for optimizing operating conditions under NAP environments.

Building upon this electron-optical foundation, detector optimization enables stable imaging at pressures up to 20,000 Pa, while localized laser-confined heating and thermionic-electron suppression extend high-temperature imaging to 1,400 °C under 6,000 Pa of air. The integrated gas-sampling architecture further enables synchronized SEM–QMS operando characterization. Representative experiments using high-temperature precursor systems, hydrated biological specimens, and catalytic reactions demonstrate the operation of the platform across these complementary environmental regimes.

Rather than introducing a series of independent instrumental components, this work establishes a unified electron-optical architecture that systematically expands the operating window of ESEM. The design principles presented here provide a general instrumentation framework for future environmental SEM systems and create new opportunities for quantitative operando investigations in catalysis, energy materials, geological processes, and biological systems under realistic reaction environments. Future developments may further improve beam utilization under higher-pressure conditions, accelerate local gas exchange, and enable increasingly quantitative correlative operando characterization across more complex environmental systems.